\documentstyle [12pt] {article}
\title{ \Large \bf Binary Black Holes in Stationary Orbits}
\author
{Sandip K. Chakrabarti\\
Tata Institute of Fundamental Research, Homi Bhabha Road, Colaba,\\
Bombay, 400005, INDIA\\}
\begin{document}
\baselineskip 22pt
\maketitle

\begin{abstract}
We show that under certain astrophysical conditions a
binary system consisting of two compact objects can be stabilized against
indefinite shrinking of orbits due to the emission of gravitational radiation.
In this case, the lighter binary companion settles down to a stable orbit when
the loss of the angular momentum due to gravitational radiation becomes equal
to its gain from the accreting matter from the disk around the more massive
primary. We claim that such systems can be stable against small perturbations
and can be regarded as steady emitters of gravitational waves of constant
frequency and amplitude. Furthermore, X-rays emitted by the secondary
can also produce astrophysically interesting situations when coupled with
gravitational lensing and Doppler effects.
\end{abstract}

\vspace{2.0cm}

It is well known that a binary system is unstable against
gravitational radiation. The radiation carries away both energy
as well as angular momentum and the orbits shrink indefinitely.
In the case of active galaxies and Quasars, it is possible to envisage
a situation, such that there is a large accretion disk
($\sim 10^6$ Schwarzschild radius)
around the central black hole (of mass $M_1\sim 10^{6-9}M_\odot$)
and a large number of smaller objects (of mass $M_2\sim 0.1-100 M_\odot$)
such as ordinary stars, white dwarfs, neutron stars and black holes
orbiting around it at various orbits.

Let us concentrate on one such system, binary in nature, where the
secondary companion is also compact--- presumably, a neutron star or a
black hole. Since a companion
in an eccentric orbit will loose angular momentum rather rapidly,
it is safe to assume that the orbit is circular. We also assume
that the secondary is in the same plane as the accretion disk
itself. This scenario is general enough, since the secondary, in any other
plane would be brought to the equatorial plane due to the Lens-Thirring
effect. The shrinking orbit of the binary will eventually bring it close
to the central massive body, inside the denser part of the accretion disk.

In the case of AGNs and Quasars, the angular momentum
of the surrounding accretion disk close to the hole could be almost constant
($l\sim 4GM_1/c$) giving rise to thick accretion disks. In between the inner
edge of such a disk (around $2r_s$) and the center of the disk (which could
be anywhere between the marginally stable orbit at $3r_s$ and, say, $10-50r_s$
depending upon the angular momentum distribution) such a disk is
super-Keplerian.
(see, e.g., Paczynski and Wiita, 1980; here $r_s=2GM_1/c^2$ is the
Schwarzschild black hole radius, $G$ and $c$ are the gravitational constant
and the velocity of light respectively). That radiation pressure could
cause such a change in the angular momentum distribution has been
well demonstrated by Maraschi, Reina and Treves (1976).
In the absence of a rigorous understanding of the viscous processes which
determine the exact distribution of angular momentum inside the disk, it is
customary to use power law distribution of angular momentum (e.g. Paczynski
and Wiita, 1980; Chakrabarti 1985). Fig. 1 shows the comparison of a few
typical angular momentum distributions as functions of the Schwarzschild
radius of the primary. Here, $l_k$ denotes the Keplerian distribution,
with a minimum at the marginally stable orbit $r_{ms}=3r_s$; $l_p$ is a
power law distribution $l_p=2.8 (r/r_s)^{0.25}$ and $l_{mb}$ is the marginally
bound angular momentum (=$4GM_1/c$). The power law distribution intersects
the Keplerian curve at two points: at $r_i$, the inner edge ($i$)
of the disk, and at $r_c$, the center ($c$) of the disk.
Therefore, once the orbiting black hole or a neutron star (which is on a
Keplerian orbit) is inside the region between $i$ and $c$,
it will accrete mass as well as {\it angular momentum}, from the disk matter.
We show below that it is possible to have a situation where the
orbital angular momentum gained by the binary companion balances exactly
the loss due to gravitational waves and the binary attains
a stable orbit, emitting steadily both the gravitational radiation
as well as X-rays and gamma rays with possible modulations due to
redshifts and Doppler shifts due to its orbital motion, as well as
gravitational lensing due to the presence of the massive primary.

We like to mention here that a similar consideration of replenishment
of angular momentum through accretion onto a rapidly rotating
neutron star radiating angular momentum through gravitational
waves was considered by Wagoner (1984). He finds that it is possible to
stabilize the neutron star from any further loss of angular momentum
and energy.

The rate of loss of energy $dE/dt$ in a binary system consisting
of two point masses $M_1$ and $M_2$ in circular orbits, having orbital
period $P$ (in hours) is given by (see, Peters and Matthews, 1963; Lang, 1980),
$$
\frac{dE}{dt}=3 \times 10^{33} (\frac {\mu}{M_\odot})^2
(\frac{M}{M_\odot})^{4/3} (\frac{P}{1 hr})^{-{10}/{3}} {\rm ergs\ sec^{-1}},
\eqno{(1)}
$$
where,
$$
\mu=\frac {M_1 M_2}{M_1+M_2}
$$
and
$$
M=M_1+M_2.
$$
The orbital angular momentum loss rate would be,
$$
\frac{dL_-}{dt}=\frac{1}{\Omega} \frac{dE}{dt}
\eqno{(2)}
$$
where $\Omega=\sqrt{G M_1/r^3}$ is the Keplerian angular velocity of the
secondary black hole orbiting at a mean radius of $r$. The negative sign
signifies a loss.

For the problem in hand, we assume that $M_1> M_2$. In fact, we shall be
typically interested in the cases, where, the primary has
$M_1 \sim 10^{6-8} M_\odot$ and the companion has $M_2 \sim 10^{0-2}M_\odot$.
We assume further that the primary is accreting at the rate
at least close to its critical rate, $\frac{{\dot M}_{1,Ed}}{\eta}$,
where ${\dot M}_{1,Ed}$ is the Eddington rate of the primary and ${\eta}$ is
the
release of the binding energy loss by the accreting matter, typically $6$
per cent, in case the primary is a Schwarzschild black hole. Thus,
disk can supply matter, several times the Eddington rate to the secondary
as it orbits around the central hole.

Let us assume that the secondary is accreting ${\dot M}_2$ amount of
matter from the disk per second. If
it accretes specific angular momentum at a fraction $\alpha <1.0$
of the marginally bound value $4G M_1/c$, the rate of gain of
of angular momentum by the companion is given by,
$$
\frac{dL_+}{dt}=\frac{4GM_1}{c} \alpha {\dot M}.
\eqno{(3)}
$$
Here $\alpha$ is assumed to be a constant  for simplicity, even though
its value is necessarily becomes zero at the inner edge $r_i$ as well as
at the center $r_c$ of the disk. It is negative for $r> r_c$.

Equating (2) and (3) we obtain the equilibrium radius as (Chakrabarti 1992),
$$
r_{eq}=15.63 \frac{{M_1}^{1/21}}{[\alpha {\dot M}_2]^{2/7}}
(\frac{\mu}{M_\odot})^{4/7} (\frac{M}{M_\odot})^{8/21}
(\frac{10^8M_\odot}{M_1})
\eqno{(4)}
$$
in units of the Schwarzschild radius of the primary. For an illustration,
we choose $M_1=10^8 M_\odot, \alpha=0.05$ and ${\dot M}_2$
as the critical rate onto the secondary. If we choose, $M_2=1M_\odot$,
we obtain, $r_{eq}=23 r_s$.

It is to be noted that in want of a rigorous distribution of
angular momentum in the disk closer to the hole, we have chosen
the rate of accretion of angular momentum to be constant
(eqn. 3), though globally it should vary significantly.
As is obvious from Fig. 1, the difference
of angular momentum which could be transported to the orbiting black hole,
$l_p - l_K$ decreases sharply as one goes from from $r_{ms}$ to the center
($c$) of the disk at $r_c$. Thus, $\alpha$ sharply falls to zero at $r_c$
and the transport can only take place upto
the center of the disk. The ${\dot M}_2$ itself may also fall very rapidly
with radial distance due to the sudden increase in thickness of the disk
close to the center, causing lesser matter to be accreted onto the secondary.
For $r> r_c$, the transport acts in the opposite
direction causing the secondary object to shrink faster than the rate
prescribed by the loss of the gravitational waves. This effect need not be
very strong, since the density of matter falls off sharply with distance
and therefore the actual accretion of matter onto
the secondary with sub-Keplerian angular momentum will be smaller.

It is easy to see that the equilibrium orbit at $r_{eq}$ is stable as long
as $r_{mb} < r_{eq} < r_c$ and the product $\alpha {\dot M}_2$ locally
falls off faster than $r^{-\frac{7}{2}}$.
That the second condition is attainable could be seen in the following
way. For simplicity, let us consider an wedge shaped disk, where the
density $\rho (r) \propto r^{-3/2}$. Let the {\it accreting} angular momentum
onto the black hole
be of the form $\alpha (r) \propto r^{-\beta}$, where $\beta$ is a constant.
The rate of accretion onto the secondary ${\dot M}_2 (r) \propto \rho (r)
v (r) \propto r^{-{5/2+\beta}}$, where $v(r)=\alpha(r)/r$ is
the relative velocity
between the disk matter and the black hole. Therefore, the product
$\alpha {\dot M}_2 \propto r^{-{5/2+2\beta}}$ could be faster
than $r^{-7/2}$ when $\beta >1/2$. In this case, as the secondary black hole is
perturbed to a higher radius, it accretes $lesser$ angular momentum from the
disk and it comes back to $r_{eq}$ through gravitational radius.
When it is pushed closer to the black hole it gains more angular momentum
than it losses, and therefore returns back to $r_{eq}$. It is possible
that, in reality, $r_{eq}$ would settle down closer to the center of the disk.
If, $r_{eq} < r_{ms}=3r_s$, the accretion of angular momentum increases
with radial distance. This makes the equilibrium orbits in this region
unstable under small perturbations.

Being on a stationary orbit, such a source will be emitting
steady gravitational waves with frequency,
$$
f=\frac{1}{\pi}(\frac {GM}{{r_{eq}}^3})^{1/2}
\eqno{(5)}
$$
and the constant amplitude of metric perturbation on earth caused by such
a source would be (e.g., Sathyaprakash and Dhurandhar, 1991)
$$
|A| =5.765 \times 10^{-19} \frac{\mu}{1 M_\odot}
(\frac {M}{10^6 M_\odot})^{2/3} f^{2/3}(\frac{D}{1 Mpc})^{-1}
\eqno{(6)}
$$
$D$ being the distance of the source from earth.
Though the frequency is much lower ($10^{-6}$ to $0.1$Hz) than the present
observable limits ($10-100$Hz), the steady source can cause periodic
variation of the distance between a pulsar and the earth which is reflected
on the the post-fit residual. Such variation is observed in the case of
PSR1937+21 (J. Taylor, private communication).

We have demonstrated that under certain astrophysical conditions, when the
accretion disk around a massive black hole is super-Keplerian, the orbit
of a lighter compact object, such as a black hole, neutron star or a white
dwarf can be stabilized against the loss of angular momentum due to
gravitational wave. Such a system can be a steady source of gravitational
waves. The accreting secondary would also emit X-rays and gamma rays which
may be modulated due to red-shifts, Doppler effects and gravitational bending
of light. Indeed, NGC6814, a low luminosity syfert galaxy, which shows
steady periodicity of X-ray flaring over at least five years,
could be easily explained by assuming a white dwarf in a stationary orbit
at $6r_s$ around a black hole (Chakrabarti and Bao, 1992).
Though we concentrated our discussion in the context of thick
accretion disks, it is easy to see that even for geometrically thin
disks stabilization of orbits may be possible, provided the disk remains
super-Keplerian and the stability conditions are satisfied.
In cases where stabilization is not possible, particularly when the
disk is not sufficiently super-Keplerian, or, when the component
masses are comparable
so that not enough mass accretion takes place onto the secondary,
the behavior of the gravitational wave emitted should still be greatly
influenced by the presence of the disk. Work along this line is in
progress and will be reported else where.

\newpage

\centerline {\bf References}

\noindent Chakrabarti, S.K., 1985, ApJ, 288, 1.\\
\noindent Chakrabarti, S.K., 1992, {\it Presented at the 13th International
Conference on the General Relativity and Gravitation}, Cordoba, Argentina.\\
\noindent Chakrabarti, S.K. and Bao, G. 1992, ICTP Preprint No. 115/92
\noindent Lang, K.R., 1980, Astrophysical Formula (Springer Verlag:New York)\\
\noindent Maraschi, L., Reina, C., and Treves, A. 1976, ApJ, 206, 295.\\
\noindent Paczynski, B. and Wiita, P.J., 1980, A \& A, 88, 23.\\
\noindent Peters, P.C. and Matthews, J., 1963, Phys. Rev, 131, 435.\\
\noindent Sathyaprakash, B.S. and Dhurandhar, S., 1991, Phys. Rev. D, 44,
3819.\\
\noindent Wagoner, R., 1984, ApJ, 378, 345.\\

\newpage

\centerline {Figure Caption}

\noindent Fig. 1: The angular momentum distribution ($l_p$) inside a typical
radiation pressure dominated disk is compared with the Keplerian
distribution $l_k$ and with the marginally bound value $l_{mb}$. The inner
edge of the disk is located at $i$ and the center is located at $c$.
An orbiting black hole between $i$ and $c$ accretes matter as well as
angular momentum from the disk.
\end{document}